\documentclass[a4paper]{jpconf}

\usepackage{amsmath}
\usepackage{xcolor}

\newcommand{\nc}{\newcommand}

\nc{\bea}{\begin{eqnarray}}  
\nc{\eea}{\end{eqnarray}}

\renewcommand{\vec}[1]{\boldsymbol{#1}}

\begin{document}

\title{The CP-symmetries of the 2HDM}
\author{B. Grzadkowski$^1$, O.M. Ogreid$^2$ and P. Osland$^3$}
\address{$^1$ Faculty of Physics, University of Warsaw, Pasteura 5, 02-093 Warsaw, Poland}
\address{$^2$ Western Norway University of Applied Sciences,
	Postboks 7030, N-5020 Bergen, Norway}
\address{$^3$ Department of Physics,
	University of Bergen, Postboks 7803, N-5020 Bergen, Norway}
\ead{bohdan.grzadkowski@fuw.edu.pl, omo@hvl.no, per.osland@uib.no}
\begin{abstract}
We discuss the three different classes of CP-symmetries that can be realized in a two-Higgs-doublet model, CP1, CP2 and CP3. We express conditions for realizing these symmetries in terms of masses and couplings of the model, thereby providing a way of verifying which, if any, of these symmetries is realized by nature.
\end{abstract}
\section{Introduction}
Multi-Higgs-doublet models provide attractive extensions of the Standard Model (SM) due to their ability to accommodate Dark Matter and/or CP-violation \cite{Branco:2011iw}. The two-Higgs-doublet model (2HDM) is the simplest of these extensions. Its CP-properties are well known and have been studied and expressed in many different ways, using different formalisms \cite{Lavoura:1994fv,Botella:1994cs,Branco:2005em,Gunion:2005ja,Ivanov:2005hg,Nishi:2006tg,Maniatis:2007vn,Grzadkowski:2014ada,Grzadkowski:2016szj}. In the 2HDM Lagrangian, interactions involving scalars are expressed using the two doublets $\Phi_1$ and $\Phi_2$. It is also possible to form combinations of these as $\Phi_i\rightarrow U_{ij}\Phi_j^\prime$, where $U_{ij}$ is any $U(2)$-matrix, and express the Lagrangian in terms of the transformed doublets. This is referred to as a change of basis. Clearly, physics should not depend upon our choice of basis, and this motivates basis-invariant formulations of physical conditions. In our formalism, we focus on expressing the CP-properties in terms of physical observable quantities only, i.e. only masses and physical couplings of the model. All masses and measurable couplings involving the scalars are naturally basis-invariant. For this purpose we introduce the physical parameter set
\bea
{\cal P}\equiv\{M_{H^\pm}^2,M_1^2,M_2^2,M_3^2,e_1,e_2,e_3,q_1,q_2,q_3,q\},
\eea
containing the squared masses of the charged scalar as well as those of the tree neutral ones. Furthermore, $e_i$ parametrizes the $H_iVV$ couplings $V=(W,Z)$, $q_i$ parametrizes the $H_iH^+H^-$ couplings and $q$ parametrizes the $H^+H^+H^-H^-$ coupling. See \cite{Grzadkowski:2014ada,Grzadkowski:2016szj,Ogreid:2018bjq,Grzadkowski:2018ohf} for more details. All physical observables arising from the bosonic sector of the 2HDM can be expressed in terms of the 11 parameters of ${\cal P}$. In order to make the translation from the parameters of the potential and the vacuum to the physical parameter set we can either translate explicitly basis-invariant quantities,
or we can translate a combination of non-invariant constraints that together constitutes a basis-invariant constraint. The method for translating from parameters of the potential to the physical parameter set is presented in detail in \cite{Ogreid:2018bjq}.
\section{Three classes of CP-symmetries}
In the 2HDM, the potential is CP-invariant provided it is symmetric under the following transformation of the Higgs-doublets,
\bea
\Phi_i\rightarrow X_{ij}\Phi_j^*,
\eea
where $X_{ij}$ is any $U(2)$-matrix.
In \cite{Ferreira:2009wh}, three different classes of CP-symmetries were presented, according to the form of the $X_{ij}$. These three symmetries were denoted CP1, CP2 and CP3;
\begin{alignat}{2}
&\!\!\bullet &\ \
&\text{\bf CP1: } X_{ij}=
\begin{pmatrix}
1 & 0\\
0 & 1
\end{pmatrix}. \nonumber \\
&\!\!\bullet &\ \
&\text{\bf CP2: }  X_{ij}=
\begin{pmatrix}
0 & 1\\
-1 & 0
\end{pmatrix}. \nonumber \\
&\!\!\bullet &\ \
&\text{\bf CP3: }  X_{ij}=
\begin{pmatrix}
\cos\theta & \sin\theta\\
-\sin\theta & \cos\theta
\end{pmatrix},\, 0<\theta<\pi/2. \nonumber
\end{alignat} 
CP1 is the ``ordinary'' CP-symmetry of the potential, which can be shown to be equivalent to the possibility of simultaneously having both a real potential and a real vacuum \cite{Gunion:2005ja}.  CP2 and CP3 denote even higher degrees of symmetry in the potential. Whenever the potential possesses the CP2 or CP3 symmetry, it also possesses the CP1 symmetry.  Basis-independent formulations of conditions for invariance of the 2HDM-potential under CP1, CP2 and CP3 were presented in \cite{Ferreira:2010yh} in terms of properties of vectors $\vec{\xi}$, $\vec{\eta}$ and a matrix $E$. For a detailed study of vacua within models invariant under these symmetries see \cite{Battye:2011jj}.
\section{Physical conditions for CP1, CP2 and CP3}
The conditions for CP1 in terms of the physical parameter set ${\cal P}$ are well known from previous works \cite{Grzadkowski:2014ada,Grzadkowski:2016szj}. In the following three sub-sections we are going to express the conditions for CP2 and CP3 derived in \cite{Ferreira:2010yh} in terms of physical parameters using the technique outlined in \cite{Ogreid:2018bjq}. Hereafter we will refer to CP1 either as CP1 or just CP.
\subsection{Physical conditions for CP1}
Here, we just repeat the results found in \cite{Grzadkowski:2014ada,Grzadkowski:2016szj}. The ``ordinary'' CP-symmetry, CP1, can be realized with four different physical configurations.
\begin{alignat}{2}
&\!\!\bullet &\ \
&\text{\bf Case A: }  M_1=M_2=M_3. \text{ (Full mass degeneracy.)} \nonumber \\
&\!\!\bullet &\ \
&\text{\bf Case B: }  M_i=M_j,\, (e_jq_i-e_iq_j)=0. \text{ (Partial mass degeneracy.)} \nonumber \\
&\!\!\bullet &\ \
&\text{\bf Case C: }  e_k=q_k=0. \text{ (No mass degeneracy.)}\nonumber\\
&\!\!\bullet &\ \
&\text{\bf Case D: }  M_{H^\pm}^2=\frac{v^2}{2D}
\left[
e_1q_1M_2^2M_3^2+e_2q_2M_1^2M_3^2+e_3q_3M_1^2M_2^2-M_1^2M_2^2M_3^2
\right], \nonumber\\
& & &\phantom{\text{\bf Case D: }}
q=\frac{1}{2D}
\left[
(e_2q_3-e_3q_2)^2M_1^2+(e_3q_1-e_1q_3)^2M_2^2+(e_1q_2-e_2q_1)^2M_3^2+M_1^2M_2^2M_3^2
\right].\nonumber
\end{alignat} 
where $D=e_1^2M_2^2M_3^2+e_2^2M_1^2M_3^2+e_3^2M_1^2M_2^2$.	
The conditions for cases A, B and C guarantee that both the potential and the vacuum conserve CP, while case D only guarantees a CP conserving potential and will lead to spontaneous CP-violation (SCPV) unless the constraints of either Case A, B or C are satisfied in addition to the condition of Case D. We will hereafter refer to Case D as the SCPV-condition.

In \cite{Grzadkowski:2014ada} we identified several physical processes involving decaying $Z$-bosons where the amplitude contains CP-odd form factors, thereby facilitating CP-violating processes. In \cite{Grzadkowski:2016lpv} we studied the effects of one-loop generated effective $ZZZ$ and $ZWW$ vertices along with a set of CP-sensitive asymmetries for $ZZ$ and $W^+W^-$ production at $e^+e^-$colliders that directly depends on a CP-odd basis-invariant.

\subsection{Physical conditions for CP2}
A CP2-symmetric 2HDM will necessarily have both a CP-conserving potential and vacuum, so the physical conditions for CP2 will be sub-cases of Cases A, B and C above. We find that the potential is CP2-invariant in the following physical configurations:
\begin{alignat}{3}
&\!\!\bullet &\ \
&\text{\bf Case ABBBD: }&&  M_1=M_2=M_3,\quad (e_2q_1-e_1q_2)=(e_3q_1-e_1q_3)=( e_2q_3-e_3q_2)=0,\nonumber\\
& & &\phantom{\text{\bf Case ABBBD: }}&&
M_{H^\pm}^2=\frac{1}{2}(e_1q_1+e_2q_2+e_3q_3-M_1^2),\quad q=\frac{M_1^2}{2v^2}. \nonumber \\
&\!\!\bullet &\ \
&\text{\bf Case BCD: }&&  M_i=M_j,\quad (e_jq_i-e_iq_j)=0,\quad e_k=q_k=0,\nonumber\\
& & &\phantom{\text{\bf Case BCD: }}&&
M_{H^\pm}^2=\frac{1}{2}(e_iq_i+e_jq_j-M_i^2),\quad q=\frac{M_i^2}{2v^2}. \nonumber \\
&\!\!\bullet &\ \
&\text{\bf Case CCD: }&&  e_j=e_k=q_j=q_k=0,\quad M_{H^\pm}^2=\frac{1}{2}(e_iq_i-M_i^2),\quad q=\frac{M_i^2}{2v^2}. \nonumber
\end{alignat} 
The logic behind this labeling is that the physical configurations present here can be interpreted as sub-cases of multiple of the four different cases (A, B, C and D) of CP1. A repeated letter indicates that two (or more) separate constraints (with different indices) of said case is needed to get CP2.
In each of these physical configurations, the SCPV-condition (Case D) is present as constraints on $M_{H^\pm}^2$ and $q$ in combination with at least two separate conditions for CP-invariance (Cases A, B or C) in addition.
In some crude sense we can say that these conditions (symmetries) are stacked on top of each other.
\subsection{Physical conditions for CP3}
A CP3-symmetric 2HDM will also necessarily have both a CP-conserving potential and vacuum, so the physical conditions for CP3 will be sub-cases of the cases for CP1. More than this, any model with CP3 symmetry will also have the CP2 symmetry. Thus, the cases of CP3 will be sub-cases of the cases we found for CP2. We find that the potential is CP3-invariant in the following physical configurations:
\begin{alignat}{3}
&\!\!\bullet &\ \
&\text{\bf Case ABBBD: }&&  \text{As in the previous sub-section}. \nonumber \\
&\!\!\bullet &\ \
&\text{\bf Case BC$_0$D: }&&  \text{As Case BCD with the additional constraint } M_k=0. \nonumber
 \\
&\!\!\bullet &\ \
&\text{\bf Case B$_0$CD: }&& \text{As Case BCD with the additional constraint } M_i=0. \nonumber
\\
&\!\!\bullet &\ \
&\text{\bf Case BCCD: } && \text{As Case CCD with the additional constraint } M_j=M_k. \nonumber \\
&\!\!\bullet &\ \
&\text{\bf Case CC$_0$D: } && \text{As Case CCD with the additional constraint } M_j=0\text{ (or }M_k=0\text{)}. \nonumber
\end{alignat} 
The subscript 0 attached to a constraint tells us that the mass corresponding to the scalar(s) involved in the constraint vanishes\footnote{If we have labeling {\bf B$_0$}, where {\bf B} refers to the constraints $M_i=M_j,\ e_jq_i-e_iq_j=0$, then the subscript 0 tells us that $M_i=M_j=0$. If we have labeling {\bf C$_0$}, where {\bf C} is the constraint $e_k=q_k=0$, then the subscript 0 tells us that $M_k=0$.}.   
Comparing to the conditions for CP2, we see that the conditions for CP3 are identical with the exception that in some cases we need additional mass degeneracy, or some mass(es) must vanish in order to get CP3.

\section{Summary}
We have expressed conditions for the CP1, CP2 and CP3 symmetries of the 2HDM in terms of observable parameters, i.e. masses and physical couplings. Experimentally, we know that if nature has in fact realized a 2HDM in which the Higgs-boson discovered at the LHC is one of the three neutral scalars of the model, then we are close to the so-called alignment limit defined by the requirement that one of the three neutral scalars ($H_1$) exactly mimics the SM-Higgs boson. In our formalism, alignment can be expressed as $e_1=v,\, e_2=e_3=0$. We would like to stress the fact that in a CP2-symmetric model, alignment is a natural consequence of the symmetry provided the neutral masses are non-degenerate. For a discussion of natural alignment in the 2HDM see \cite{Dev:2014yca}. The conditions for the CP1, CP2 and CP3 symmetries of the 2HDM presented here provide a direct way to verify if nature has realized these symmetries in the 2HDM.


\section*{References}

\begin{thebibliography}{100}
\bibitem{Branco:2011iw}
Branco G C, Ferreira P M, Lavoura L, Rebelo M N, Sher M and Silva J P,
{\it Phys. Rept. }  {\bf 516} (2012) 1

\bibitem{Lavoura:1994fv}
Lavoura L and Silva J P,
{\it Phys. Rev. }D {\bf 50} (1994) 4619

\bibitem{Botella:1994cs}
Botella F J and Silva J P,
{\it Phys. Rev. }D {\bf 51} (1995) 3870

\bibitem{Branco:2005em}
Branco G C, Rebelo M N and Silva-Marcos J I,
{\it Phys. Lett. }B {\bf 614} (2005) 187

\bibitem{Gunion:2005ja}
Gunion J F and Haber H E,
{\it Phys. Rev. }D {\bf 72} (2005) 095002

\bibitem{Ivanov:2005hg}
Ivanov I P,
{\it Phys. Lett. }B {\bf 632} (2006) 360

\bibitem{Nishi:2006tg}
Nishi C C,
{\it Phys. Rev. }D {\bf 74} (2006) 036003
Erratum: [{\it Phys. Rev.} D {\bf 76} (2007) 119901]

\bibitem{Maniatis:2007vn}
Maniatis M, von Manteuffel A and Nachtmann O,
{\it Eur. Phys. J. }C {\bf 57} (2008) 719

\bibitem{Grzadkowski:2014ada}
Grzadkowski B, Ogreid O M and Osland P,
{\it JHEP }{\bf 1411} (2014) 084

\bibitem{Grzadkowski:2016szj}
Grzadkowski B, Ogreid O M and Osland P,
{\it Phys. Rev. }D {\bf 94} (2016) no.11,  115002

\bibitem{Ogreid:2018bjq}
Ogreid O M,
{\it PoS CORFU }{\bf 2017} (2018) 065

\bibitem{Grzadkowski:2018ohf}
Grzadkowski B, Haber H E, Ogreid O M and Osland P,
{\it JHEP }{\bf 1812} (2018) 056

\bibitem{Ferreira:2009wh}
Ferreira P M, Haber H E and Silva J P,
{\it Phys. Rev. }D {\bf 79} (2009) 116004

\bibitem{Ferreira:2010yh}
Ferreira P M, Haber H E, Maniatis M, Nachtmann O and Silva J P,
{\it Int. J. Mod. Phys. }A {\bf 26} (2011) 769

\bibitem{Battye:2011jj}
Battye R A, Brawn G D and Pilaftsis A,
{\it JHEP }{\bf 1108} (2011) 020

\bibitem{Grzadkowski:2016lpv}
Grzadkowski B, Ogreid O M and Osland P,
{\it JHEP} {\bf 1605} (2016) 025
Erratum: [{\it JHEP} {\bf 1711} (2017) 002]

\bibitem{Dev:2014yca}
Dev P S B and Pilaftsis A,
{\it JHEP} {\bf 1412} (2014) 024
Erratum: [{\it JHEP} {\bf 1511} (2015) 147]

\end{thebibliography}
\end{document}